\documentclass[letterpaper, 10 pt, conference]{ieeeconf}
\IEEEoverridecommandlockouts          
\usepackage{algorithm}
\usepackage{comment}
\usepackage{setspace}
\usepackage{float}
\let\labelindent\relax
\usepackage{enumitem}
\usepackage[fancythm,fancybb,morse,elsart]{jphmacros2e}  
\usepackage{algpseudocode}
\usepackage{dsfont}
\usepackage{stackengine}
\usepackage{amssymb}
\usepackage{mathtools} 
\usepackage{tikz}
\usepackage{comment}
\usepackage{amsmath}
\usepackage{booktabs}
\usepackage{bigints}
\usepackage{textcomp}
\usepackage{stackengine}
\usepackage{siunitx}
\usepackage{balance}
\usepackage{graphicx}          
\usepackage{xcolor}  
\usepackage[normalem]{ulem}
\usepackage{hyperref}
\usepackage{tabularx}
\usepackage{collcell}
\usepackage{cite}
\usepackage{bbm}
 \usepackage[norefs,nocites]{refcheck}
\usepackage{graphicx}
\usepackage{resizegather}

\allowdisplaybreaks

\makeatletter
\renewcommand*\env@matrix[1][\arraystretch]{%
  \edef\arraystretch{#1}%
  \hskip -\arraycolsep
  \let\@ifnextchar\new@ifnextchar
  \array{*\c@MaxMatrixCols c}}
\makeatother

\title{\textbf{RA-PI-DeepONet: A Physics-Informed Neural Operator for Reach–Avoid Differential Games}}

\author{Nick-Marios T. Kokolakis, Alan John Varghese, Shanqing Liu, Paris Perdikaris,\\ Jérôme Darbon, Rahul Mangharam, and George E. Karniadakis%
\thanks{N.-M.~T.~Kokolakis, P.~Perdikaris, and R.~Mangharam are with the University of Pennsylvania, Philadelphia, PA $19104$, USA. E-mail: \texttt{\{nmkoko, pgp, rahulm\}@engineering.upenn.edu}.}%
\thanks{A.~J.~Varghese, S.~Liu, J.~Darbon, and G.~E.~Karniadakis are with Brown University, Providence, RI $02912$, USA. E-mail: \texttt{\{alan\_john\_varghese, shanqing\_liu, jerome\_darbon, george\_karniadakis\}@brown.edu}.}}

\begin{document}
\maketitle
\thispagestyle{empty}
\pagestyle{empty}
\begin{abstract}
Reach-avoid differential games provide a game-theoretic control framework for certifying safety- and time-critical task completion under adversarial disturbances. However, existing solvers are computationally intensive, often suffer from the curse of dimensionality, and address only a fixed safety constraint while requiring rerunning whenever the constraint changes. In this paper, we develop a physics-informed neural operator architecture, termed RA-PI-DeepONet, for approximating the reach-avoid solution operator without requiring paired input-output training data. Specifically, we formulate the reach-avoid problem as a two-player zero-sum differential game whose value function is the unique viscosity solution of a Hamilton-Jacobi-Isaacs (HJI) variational inequality subject to a boundary condition involving an avoid function that encodes the safety constraint. We define the solution operator as a mapping from an avoid function to the corresponding value function and prove that this operator is Lipschitz continuous with respect to the supremum norm. Since the solution operator is nonlinear and implicit, evaluating the operator for a given avoid function requires approximating the corresponding solution of the analytically intractable HJI variational inequality. By the universal approximation theorem for operators and the Lipschitz continuity of the solution operator, RA-PI-DeepONet can approximate the solution operator to arbitrary accuracy on compact sets of avoid functions. The training of RA-PI-DeepONet is self-supervised, using the HJI variational inequality residual as the loss function. Once trained, RA-PI-DeepONet enables real-time value-function approximation for unseen avoid functions without re-solving the HJI variational inequality, while enforcing the boundary condition by design. Finally, numerical examples demonstrate the efficacy of the proposed framework.
\end{abstract}

\section{Introduction}
\textit{Trustworthy autonomy} requires strategic decision-making mechanisms that ensure safe and timely task completion in dynamic and adversarial environments. To this end, autonomous systems need adaptive game-theoretic safety verification frameworks that rapidly account for changes in the environment. For instance, an autonomous vehicle may need to update its safety certificates online as obstacles appear, move, or change size. This need motivates the integration of \textit{Hamilton-Jacobi reachability analysis}~\cite{mitchell2005time,bansal2017hamilton,chen2018hamilton} and \textit{scientific machine learning}~\cite{cuomo2022scientific} to develop scalable and generalizable physics-informed learning architectures that provide real-time safety assurances in changing operating environments.

\textit{Reach-avoid control problems} have been extensively studied in control theory and safety verification as a framework for characterizing the set of initial states from which there exists an admissible controller that steers the system state to a target set in finite time while avoiding an unsafe set~\cite{bokanowski2010reachability,lee2020hopf}. However, these problems do not account for adversarial disturbances.
A \textit{reach-avoid differential game} is a game-theoretic control framework for certifying safety- and time-critical task completion under adversarial disturbances~\cite{margellos2011hamilton,fisac2015reach}. Specifically, a reach-avoid differential game is a two-player zero-sum differential game wherein the controller seeks to steer the system trajectory into a target set within a prescribed time horizon while satisfying state constraints, whereas the adversary seeks to prevent the controller from satisfying these requirements. A \textit{reach-avoid set} comprises the initial states from which the controller can satisfy these requirements despite the adversary and is characterized as the zero sublevel set of the value function, which is the unique viscosity solution of a Hamilton-Jacobi-Isaacs (HJI) variational inequality. However, the HJI variational inequality is generally analytically intractable; hence, \textit{level set methods}~\cite{mitchell2002application} are typically used to numerically approximate the value function. Nevertheless, these grid-based numerical HJI solvers are computationally intensive and suffer from the curse of dimensionality.

To mitigate the computational limitations of grid-based level set methods, \textit{physics-informed learning} has emerged as a promising approach for developing mesh-free neural PDE solvers~\cite{karniadakis2021physics}. Specifically, \textit{physics-informed neural networks} (PINNs) are self-supervised deep learning frameworks that incorporate the governing partial differential equation and boundary conditions into the training loss, thereby enabling solution approximation without using input-output pairs generated by numerical solvers~\cite{raissi2019physics}. Leveraging PINNs, \textit{DeepReach} was developed in~\cite{bansal2021deepreach} to approximate the value function associated with the reach-avoid differential game. However, DeepReach learns the value function for a fixed reach-avoid problem; when the avoid set changes, DeepReach must be retrained. 

Addressing this limitation requires characterizing the \textit{solution operator} associated with the HJI variational inequality, which maps an avoid function encoding the avoid set to the corresponding value function.
However, this solution operator is nonlinear and implicit, and hence, evaluating the operator for a given avoid function requires solving the corresponding HJI variational inequality. Thus, learning the solution operator over a class of avoid functions is essential.

The \textit{Deep Operator Network} (DeepONet) is a neural operator architecture that, by the universal approximation theorem for operators, approximates nonlinear continuous operators on compact sets of continuous functions to arbitrary accuracy~\cite{lu2021learning}. However, DeepONet training is supervised using paired input-output training data, which are typically generated by computationally expensive numerical PDE solvers. To address this limitation, \textit{physics-informed DeepONet} (PI-DeepONet) builds on DeepONet by incorporating the governing PDE and boundary conditions into the training loss, enabling solution-operator learning without paired input-output training data~\cite{wang2021learning}. Nevertheless, PI-DeepONet enforces boundary conditions as soft constraints through penalty terms in the training loss, which introduces additional loss-balancing hyperparameters and does not guarantee exact satisfaction of the boundary conditions. Related neural Hamilton-Jacobi reachability frameworks \textit{without state constraints} have considered supervised neural-operator learning over varying target functions~\cite{li2025hjrno}, parameter-conditioned value-function approximation~\cite{borquez2023parameter}, and exact enforcement of the terminal boundary condition~\cite{kokolakis2026fixed,singh2025exact}. To the best of our knowledge, no existing work has developed a physics-informed DeepONet framework for learning the solution operator associated with the HJI variational inequality of reach-avoid differential games while enforcing the terminal boundary condition by construction.

\paragraph*{Contributions} The contributions of this paper are threefold. First, we establish that the solution operator associated with the HJI variational inequality for a reach-avoid differential game is Lipschitz continuous with respect to the supremum norm. Second, we develop a physics-informed neural operator framework, termed \textit{RA-PI-DeepONet}, for approximating the solution operator without requiring paired input-output training data and show that RA-PI-DeepONet can approximate the solution operator to arbitrary accuracy on compact sets of avoid functions. Finally, RA-PI-DeepONet enables real-time value-function approximation for unseen avoid functions without re-solving the HJI variational inequality, while enforcing the terminal boundary condition by construction.

\paragraph*{Notation}
The notation used in this paper is standard. For a bounded function $f:\Omega\to\mathbb R$, with $\Omega\subseteq\mathbb R^n$, define the supremum norm
$
\|f\|_{\infty,\Omega}\triangleq\sup_{x\in\Omega}|f(x)|.
$
When $\Omega=\mathbb R^n$, we write $\|f\|_\infty$. For a set $X \subseteq \mathbb R^n$, $C(X)$ and $C^1(X)$ denote the sets of continuous and continuously differentiable functions on $X$, respectively. For a function $f:X\to\mathbb R$ and $Y\subseteq X$,
$f|_Y$ denotes the restriction of $f$ to $Y$.

\section{Reach-Avoid Differential Games} 
In this section, we characterize the continuity properties of the solution operator associated with a reach-avoid differential game. Specifically, we formulate the reach-avoid problem as a two-player zero-sum differential game characterized by a value function, which is the unique viscosity solution of an HJI variational inequality. We define the solution operator as a mapping from the avoid function to the corresponding value function and show that the operator is Lipschitz continuous. 

Consider the controlled nonlinear dynamical system given by
\begin{align} \label{eq: system}
\hspace{-0.14 cm}\dot{x}(t)&=F(x(t),u(t),d(t)),\quad x(t_{0})=x_{0},\quad t_{0} \leq t\leq t_{\mathrm{f}},
\end{align}
where, for every $t \in \left[t_0, t_{\mathrm{f}}\right]$, $x(t) \in \mathbb{R}^n$ is the state vector, and  $u(t) \in U \subset \mathbb{R}^{m_u}$ and $d(t) \in D \subset \mathbb{R}^{m_d}$ are the control and adversarial inputs with $U$ and $D$ being nonempty compact sets.
The mapping $F :  \mathbb{R}^n \times U \times D \to \mathbb{R}^n$
is bounded, globally Lipschitz continuous in $x$ uniformly in $(u,d)$, and jointly continuous in $(u,d)$.

The control input $u(\cdot)$ and the adversarial input $d(\cdot)$ in \eqref{eq: system} are restricted to the class of \textit{admissible control inputs} $\mathcal{U} \triangleq\left\{u: \left[t_0, t_{\mathrm{f}}\right] \rightarrow U: u(\cdot)\right.$ is Lebesgue measurable$\}$ and \textit{admissible adversarial inputs} $\mathcal{D} \triangleq\left\{d: \left[t_0, t_{\mathrm{f}}\right] \rightarrow D: d(\cdot)\right.$ is Lebesgue measurable$\}$. Under the above assumptions, the required properties for the existence and uniqueness of solutions to \eqref{eq: system} are satisfied. Let $s\left(t, t_0, x_0, u(\cdot), d(\cdot)\right),\ t_0 \leq t \leq t_{\mathrm{f}}$, denote the solution to \eqref{eq: system} with initial time $t_0$, initial condition $x_0$, admissible control input $u(\cdot) \in \mathcal{U}$, and admissible adversarial input $d(\cdot) \in \mathcal{D}$.

To specify the \textit{information structure} of the game, we assume that the controller is restricted to \textit{nonanticipative} strategies. Let $\Gamma$ be the class of nonanticipative strategies defined as $\Gamma \triangleq
\Big\{
\gamma:\mathcal D \to \mathcal U: 
\text{ for all } \tau \in [t_0,t_f] \text{ and all } d(\cdot),\hat d(\cdot)\in\mathcal D,
\text{ if } d(t)=\hat d(t) \text{ for almost every } t \in [t_0,\tau],
\text{ then } \gamma[d](t)=\gamma[\hat d](t)
\text{ for almost every } t \in [t_0,\tau]
\Big\}.$ Hence, the information structure is \textit{asymmetric}, since the controller uses nonanticipative strategies with respect to the adversarial input, while the adversary selects inputs in $\mathcal D$.

Let $r : \mathbb{R}^n \to \mathbb{R}$ be a bounded and globally Lipschitz continuous 
\textit{reach function}. The \textit{reach set} is defined by
\begin{equation} \label{eq: reach set}
R \triangleq \{ x \in \mathbb{R}^n : r(x) \leq0 \}.
\end{equation}

The class of admissible \textit{avoid functions} is defined by

\begin{align} \label{eq: set of avoid functions}
\mathcal{A}
\triangleq 
\{&
 a : \mathbb{R}^n \to \mathbb{R}: a(\cdot) \text{ is bounded and} \nonumber\\& \text{globally Lipschitz continuous}\}.
 \end{align}
Given an avoid function $a \in \mathcal{A}$, the \textit{avoid set} is defined as
\begin{equation} \label{eq: avoid set}
A\left(a\right) \triangleq \{ x \in \mathbb{R}^n : a(x) > 0 \}.
\end{equation} 
The objective of the controller is to steer the system state $x(\cdot)$ to the reach set $R$ at some time $\tau \in [t_0,t_{\mathrm{f}}]$ while ensuring $x(t) \notin A(a)$ for all $t \in [t_0,\tau]$, despite adversarial inputs. 

Given the reach set $R$, for every avoid function $a \in \mathcal{A}$, the \textit{reach-avoid set} $\mathcal{S}(R,A\left(a\right))$ of the dynamical system \eqref{eq: system} is defined as 
\begin{align*}
\mathcal S(R, A(a))
\triangleq
\{& x \in \mathbb R^n :
\text{ there exists } \gamma \in \Gamma \text{ such that,}\\
& \text{for every } d(\cdot) \in \mathcal D,
\text{ there exists } \tau \in [t_0,t_f] \\
& \text{such that }
s(\tau,t_0,x,\gamma[d](\cdot),d(\cdot)) \in R \text{ and}\\
&
s(t,t_0,x,\gamma[d](\cdot),d(\cdot)) \notin  A(a), 
\ t \in [t_0,\tau]
\},
\end{align*}
which comprises the initial states from which the reach-avoid objective is ensured.

To quantify the reach-avoid objective for the given reach set $R$ and an avoid function $a \in \mathcal{A}$, for every $ t_0 \in \mathbb{R}$, $ x_0 \in \mathbb{R}^n$, $ u(\cdot) \in \mathcal{U} $, and $ d(\cdot) \in \mathcal{D}$, define the cost functional 
\begin{align} \label{performance_measure}
J_a(t_0,x_0,u(\cdot),d(\cdot))
\triangleq
& \min_{\tau \in [t_0,t_f]}
\max\Big\{ r\left(x(\tau)\right), \nonumber \\
& \max_{\sigma \in [t_0,\tau]} a\left(x(\sigma)\right)
\Big\}.
\end{align}
The controller $u(\cdot)$ is a player that \textit{minimizes} the cost functional \eqref{performance_measure}, while the adversary $d(\cdot)$ is a player that \textit{maximizes} the cost functional \eqref{performance_measure}.

In light of the above, for every $a \in \mathcal{A}$, we cast the reach-avoid problem as a \textit{two-player zero-sum differential game}, characterized by the value function
\begin{align} \label{eq: value function}
V_a(t_0,x_0)
\triangleq
\inf_{\gamma \in \Gamma}
\sup_{d(\cdot)\in\mathcal D}
J_a\big(t_0,x_0,\gamma[d](\cdot),d(\cdot)\big),&\nonumber \\ 
(t_0,x_{0}) \in \mathbb{R} \times \mathbb{R}^{n}&,
\end{align}
subject to the dynamical system \eqref{eq: system}. 

For a fixed $a \in \mathcal{A}$, the next theorem characterizes the reach-avoid set $\mathcal{S}(R,A\left(a\right))$ as the zero sublevel set of the value function $V_a(t_0,\cdot)$. 
\begin{theorem} [\cite{margellos2011hamilton}] \label{Thm: Nash equilibrium}
Let $\mathcal{A}$ be the class of avoid functions given by \eqref{eq: set of avoid functions} and fix $a \in \mathcal{A}$. Consider the controlled nonlinear dynamical system \eqref{eq: system} with the reach set \eqref{eq: reach set}, avoid set \eqref{eq: avoid set}, and cost functional \eqref{performance_measure}. Let $V_a : \left[t_{0}, t_{\mathrm{f}}\right] \times \mathbb{R}^n \to \mathbb{R}$ be the value function given by \eqref{eq: value function}. Then, $V_a(\cdot,\cdot)$ is bounded and globally Lipschitz continuous. Moreover, $V_a(\cdot,\cdot)$ is the unique viscosity solution of the Hamilton–Jacobi–Isaacs variational inequality
\begin{align} \label{eq: HJB}
\max\Bigg\{ &
\frac{\partial V_a}{\partial t}(t,x)
 +\min\Bigg\{0,\;
\sup_{d \in  D}\inf_{u \in  U}
\frac{\partial V_a}{\partial x}(t,x)\, F(x,u,d)
\Bigg\},\nonumber\\ &  a(x) - V_a(t,x)
\Bigg\}=0,\quad (t, x) \in \left[t_{0}, t_{\mathrm{f}}\right) \times \mathbb{R}^n, 
\end{align}
subject to the terminal boundary condition
\begin{equation} \label{thm_eq:boundary condition}
V_a( t_{\mathrm{f}},x)=\max \{r(x), a(x)\},\quad x \in \mathbb{R}^{n}.
\end{equation}
Finally, the reach-avoid set $\mathcal{S}(R,A\left(a\right))$ is given by
\begin{equation*} 
 \mathcal{S}(R,A\left(a\right)) = \{x \in \mathbb{R}^n: V_a(t_{0},x) \leq 0 \}.
\end{equation*}
\end{theorem} 

Next, define the family of value functions by
\begin{equation}\label{eq: set of value functions}
\mathcal{V}
\triangleq 
\{ V_a : \left[t_{0}, t_{\mathrm{f}}\right] \times \mathbb{R}^n \to \mathbb{R}: a \in \mathcal{A}\}.
\end{equation} 
Let $\mathcal T:\mathcal A \to \mathcal V$ denote the operator defined by
\begin{equation} \label{eq:pde operator}
 \hspace{-2.8mm}  \mathcal{T}[a](t,x) \triangleq V_a(t,x),\quad a \in \mathcal{A}, \quad (t,x)\in [t_0,t_{\mathrm f}] \times \mathbb R^n.
\end{equation}
 The next theorem shows that $\mathcal {T}$ is uniformly Lipschitz continuous on $\mathcal A$.
\begin{theorem} \label{th: Lip continuity}
Let $\mathcal{A}$ be the class of avoid functions given by \eqref{eq: set of avoid functions}, let $\mathcal{V}$ be the class of value functions given by \eqref{eq: set of value functions}, and let $\mathcal T:\mathcal A \to \mathcal V$ be the operator defined by \eqref{eq:pde operator}. Then, $\mathcal T$ is uniformly Lipschitz continuous on $\mathcal A$ with Lipschitz constant $1$; that is,
\begin{equation}\label{eq:T-1-Lipschitz}
\|\mathcal T[a_1]-\mathcal T[a_2]\|_\infty
\leq
\|a_1-a_2\|_\infty, \quad  a_1, a_2 \in \mathcal A.
\end{equation}
\end{theorem}
\begin{proof}
Let $a_1, a_2\in\mathcal A$. Note that 
\begin{equation} \label{eq:a ineq}
a_1(x)\leq a_2(x)+\|a_1-a_2\|_\infty,\quad x\in\mathbb R^n.
\end{equation}
Let $\gamma \in \Gamma$, let $d(\cdot) \in \mathcal D$, and let $x(t)= s\left(t,t_0,x_0,\gamma[d](\cdot),d(\cdot)\right),\ t_0 \leq t \leq t_{\mathrm{f}}$, be the solution to \eqref{eq: system}. Then, by \eqref{eq:a ineq}, for every $\tau\in[t_0,t_f]$, we obtain 
\begin{equation} \label{eq: maxx a ineq}
\max_{\sigma \in [t_0,\tau]} a_1(x(\sigma))\leq \max_{\sigma \in [t_0,\tau]} a_2(x(\sigma))+\|a_1-a_2\|_\infty.
\end{equation}
Next, using $\max\{p,q_1\}\leq \max\{p,q_2\}$ for every $p, q_1, q_2 \in \mathbb{R}$ with $q_1 \leq q_2$, and 
$\max\{p,q_2+c\}\leq\max\{p,q_2\}+c$ for every $p, q_2 \in \mathbb{R}$ and $c\ge 0$, and letting, for every 
$\tau\in[t_0,t_f]$, 
$p=r(x(\tau))$, 
$q_1=\max_{\sigma\in[t_0,\tau]}a_1(x(\sigma))$, 
$q_2=\max_{\sigma\in[t_0,\tau]}a_2(x(\sigma))$, and $c=\|a_1-a_2\|_\infty$, it follows from \eqref{eq: maxx a ineq} that
\begin{align*}
& \max\Big\{ r\big(x(\tau)\big),\ \max_{\sigma\in[t_0,\tau]} a_1\big(x(\sigma)\big) \Big\}\\& \leq \max\Big\{ r\big(x(\tau)\big),\ \max_{\sigma\in[t_0,\tau]} a_2\big(x(\sigma)\big) \Big\}+\|a_1-a_2\|_\infty,\\ &\hspace{6cm} \tau\in[t_0,t_f],
\end{align*} 
which, by \eqref{performance_measure}, for every $ t_0 \in \mathbb{R}$, $ x_0 \in \mathbb{R}^n$, $\gamma \in \Gamma$, and $ d(\cdot) \in \mathcal{D}$, implies 
\begin{align*}
J_{a_1}\big(t_0,x_0,\gamma[d](\cdot),d(\cdot)\big) \leq J_{a_2}\big(t_0,x_0,\gamma[d](\cdot),d(\cdot)\big)\\+\|a_1-a_2\|_\infty,
\end{align*}
and hence, using \eqref{eq: value function} yields
\begin{align} \label{eq: upper ineq}
V_{a_1}(t_0,x_0)-V_{a_2}(t_0,x_0)
 \leq \|a_1-a_2\|_\infty,&\nonumber \\ (t_0,x_{0}) \in \mathbb{R} \times \mathbb{R}^{n}.&
\end{align}
A symmetric argument yields
\begin{align} \label{eq: lower ineq}
-\|a_1-a_2\|_\infty  \leq V_{a_1}(t_0,x_0)-V_{a_2}(t_0,x_0),&\nonumber \\ (t_0,x_{0}) \in \mathbb{R} \times \mathbb{R}^{n}.&
\end{align}
Now, by \eqref{eq:pde operator}, \eqref{eq: upper ineq} and \eqref{eq: lower ineq} imply \eqref{eq:T-1-Lipschitz}.\frQED 
\end{proof}
Note that $\mathcal T$ is a nonlinear \textit{implicit} solution operator associated with the HJI variational inequality \eqref{eq: HJB} subject to the boundary condition \eqref{thm_eq:boundary condition}, which is generally analytically intractable. Hence, evaluating $\mathcal T[a]$ for a given avoid function $a \in \mathcal{A}$ requires computing an approximate solution of \eqref{eq: HJB}, which is computationally intensive. To mitigate this computational burden, the next section develops a physics-informed neural operator framework to approximate $\mathcal T$ over the class of avoid functions $\mathcal A$, enabling direct value-function approximation for an \textit{unseen} avoid function without resolving the HJI variational inequality.

\section{RA-PI-DeepONet} 
In this section, we develop RA-PI-DeepONet, a physics-informed deep operator network that approximates the solution operator $\mathcal T$ without requiring precomputed value functions as training data. Specifically, we first introduce a compact class of avoid functions over which the approximation is sought. Next, we present the DeepONet architecture and a hard boundary-condition ansatz that enforces the terminal condition \eqref{thm_eq:boundary condition} by construction. We then establish the universal approximation property of RA-PI-DeepONet over this compact class. Finally, we present a physics-informed training scheme based on the residual of the HJI variational inequality and construct a neural reach-avoid set induced by the trained network.

\subsection{A Compact Class of Avoid Functions} 
Let $\Omega \subset \mathbb R^n$ be a nonempty compact set on which the value function $V_a(\cdot,\cdot)$, $a\in\mathcal A$, is approximated. However, note that the computation of $V_a(t,x)$ given by \eqref{eq: value function} at $(t,x)\in [t_0,t_{\mathrm f}]\times\Omega$ involves evaluating the reach and avoid functions along system trajectories that may leave $\Omega$. Hence, the avoid function $a(\cdot)$ is sampled over a compact set $\Omega' \supseteq \Omega$ such that every system trajectory starting in $\Omega$ at any $t\in[t_0,t_{\mathrm f}]$ remains in $\Omega'$ over the time interval $[t,t_{\mathrm f}]$. The following proposition characterizes $\Omega'$.

\begin{proposition} \label{pr: inflated set}
Consider the controlled nonlinear dynamical system \eqref{eq: system} and let $\Omega \subset \mathbb R^n$ be a nonempty compact set. Define $
c_F \triangleq \sup_{(x,u,d)\in\mathbb R^n\times U\times D}\|F(x,u,d)\|
$ and 
\begin{equation}\label{eq: inflated set}
\Omega' \triangleq \Big\{ x \in \mathbb R^n : \inf_{y \in \Omega} \|x - y\| \leq c_F \, (t_{\mathrm f} - t_0) \Big\}.
\end{equation}
Then $\Omega\subseteq\Omega'$ and $\Omega'$ is compact. Furthermore, $s\left(\tau,t,x,u(\cdot),d(\cdot)\right)\in\Omega',\ \tau\in[t,t_{\mathrm f}]$, with $t\in[t_0,t_{\mathrm f}]$, $x\in\Omega$, $u(\cdot)\in\mathcal U$, and $d(\cdot)\in\mathcal D$.
\end{proposition}
\begin{proof}
The proof is omitted for brevity.\frQED
\end{proof}
Let $\Omega'$ be given by \eqref{eq: inflated set} and let $M, L>0$. Define
\begin{align}\label{eq:compact_avoid_class}
 \mathcal A_{\mathrm c}(\Omega',M,L& )
\triangleq
\Big\{
a|_{\Omega'}: a \in \mathcal A,\quad
\|a\|_{\infty,\Omega'} \leq M,\nonumber\\
& |a(x)-a(y)| \leq L\|x-y\|,
\quad x,y\in\Omega'
\Big\}.
\end{align}
The following proposition shows that $\mathcal A_{\mathrm c}(\Omega',M,L)$ is a compact subset of $C(\Omega')$.
\begin{proposition}
Let $\Omega\subset\mathbb R^n$ be a nonempty compact set, let $\Omega'\subset\mathbb R^n$ be given by \eqref{eq: inflated set}, let $M,L>0$, and let $\mathcal A_{\mathrm c}(\Omega',M,L)$ be given by \eqref{eq:compact_avoid_class}. Then, $\mathcal A_{\mathrm c}(\Omega',M,L)$ is a compact subset of $C(\Omega')$.
\end{proposition}
\begin{proof}
The proof is omitted for brevity. \frQED
\end{proof}
Note that avoid functions constructed as truncated signed distance functions in Section~\ref{sec: simulations} are bounded and globally Lipschitz continuous, and hence, when restricted to $\Omega'$, belong to $\mathcal A_{\mathrm c}(\Omega',M,L)$ for some $M,L>0$.

For every $a_1, a_2 \in \mathcal A$ and every $x\in\Omega'$, $a_1(x)\leq a_2(x)+\|a_1-a_2\|_{\infty,\Omega'}$, and by Proposition \ref{pr: inflated set} every solution of \eqref{eq: system} with initial time in $[t_0,t_{\mathrm f}]$ and initial condition in $\Omega$ evolves in $\Omega'$. Hence, for every $(t,x)\in[t_0,t_{\mathrm f}]\times\Omega$,
repeating the proof of Theorem \ref{th: Lip continuity}
along such solutions, with \eqref{eq:a ineq} replaced
by the preceding inequality, yields
\begin{equation}\label{eq: local Lipschitz}
|\mathcal T[a_1](t,x)-\mathcal T[a_2](t,x)|
\leq \|a_1-a_2\|_{\infty,\Omega'}.
\end{equation}
    In particular, if $a_1|_{\Omega'}=a_2|_{\Omega'}$, then
    $\mathcal T[a_1]|_{[t_0,t_{\mathrm f}]\times\Omega}
    =\mathcal T[a_2]|_{[t_0,t_{\mathrm f}]\times\Omega}$. Hence, $\mathcal T$ induces a well-defined Lipschitz continuous
    operator from $\mathcal A_{\mathrm c}(\Omega',M,L)$ into
    $C([t_0,t_{\mathrm f}]\times\Omega)$, given by
    $\mathcal T[a|_{\Omega'}]
    =\mathcal T[a]|_{[t_0,t_{\mathrm f}]\times\Omega},
    \ a|_{\Omega'}\in\mathcal A_{\mathrm c}(\Omega',M,L)$. The induced operator satisfies the hypotheses of the
    generalized universal approximation theorem for operators
    \cite{lu2021learning}, motivating its approximation
    by a DeepONet.

\subsection{RA-PI-DeepONet Architecture} 
Let $m \in \mathbb N$ and let $\{x_j\}_{j=1}^{m} \subset \Omega'$ be a fixed set of \textit{sensor points}. For every $a \in \mathcal A_{\mathrm c}(\Omega',M,L)$, define the encoded representation
\begin{equation}\label{eq: encoded}
\bar a \triangleq [a(x_1),\ a(x_2),\ \ldots,\ a(x_m)]^{\mathrm T} \in \mathbb R^m.
\end{equation}
The RA-PI-DeepONet is composed of two neural networks: a \textit{branch network} $\mathcal B(\cdot;\theta_{\mathrm b}):\mathbb R^m\to\mathbb R^p$, with parameters $\theta_{\mathrm b}$, that maps the encoded representation $\bar a$ to a vector of coefficients
$
\mathcal B(\bar a;\theta_{\mathrm b}) \triangleq [\mathcal B_1(\bar a;\theta_{\mathrm b}),\ \ldots,\ \mathcal B_p(\bar a;\theta_{\mathrm b})]^{\mathrm T},
$
and a \textit{trunk network} $\mathcal N(\cdot,\cdot;\theta_{\mathrm n}): [t_0,t_{\mathrm f}]\times\mathbb R^n \to \mathbb R^p$, with parameters $\theta_{\mathrm n}$, that maps a query point $(t,x)$ to a vector of basis functions
$
\mathcal N(t,x;\theta_{\mathrm n}) \triangleq [\mathcal N_1(t,x;\theta_{\mathrm n}),\ \ldots,\ \mathcal N_p(t,x;\theta_{\mathrm n})]^{\mathrm T},
$
where $p \in \mathbb N$ denotes the latent embedding dimension.

The DeepONet approximation of $\mathcal T[a](t,x)$ on $ [t_0,t_{\mathrm f}]\times\Omega$ is given by
\begin{align} \label{eq: deeponet}
\hat{\mathcal T}_{\theta}[a](t,x) \triangleq \sum_{k=1}^p \mathcal B_k(\bar a;\theta_{\mathrm b})\, \mathcal N_k(t,x;\theta_{\mathrm n}) + b_0,& \nonumber \\ (t,x) \in [t_0,t_{\mathrm f}]\times\Omega,&
\end{align} 
where $\theta \triangleq (\theta_{\mathrm b}, \theta_{\mathrm n}, b_0)$ collects all trainable parameters and $b_0 \in \mathbb R$ is a bias term. 
The trunk network $\mathcal N(\cdot,\cdot;\theta_{\mathrm n})$ employs smooth activation functions, and hence, the map $(t,x)\mapsto\hat{\mathcal T}_\theta[a](t,x)$ belongs to $C^1([t_0,t_{\mathrm f}]\times\mathbb R^n)$ for every $a\in \mathcal A_{\mathrm c}(\Omega',M,L)$, so that the spatial and temporal derivatives required to evaluate the HJI residual can be computed by reverse-mode automatic differentiation through the trunk network.

To exactly enforce the terminal condition \eqref{thm_eq:boundary condition}, we further consider the hard boundary-condition ansatz
\begin{align} \label{eq: hard BC}
& \hat{\mathcal T}_\theta^{\mathrm{hbc}}[a](t,x) \triangleq \max\{r(x),a(x)\} + (t_{\mathrm f}-t)\, \mathcal G_\theta[a](t,x), \nonumber\\ & \hspace{4.5cm}
(t,x)\in[t_0,t_{\mathrm f}]\times\Omega.
\end{align}
where $\mathcal G_\theta[a](t,x) \triangleq \sum_{k=1}^p \mathcal B_k(\bar a;\theta_{\mathrm b})\, \mathcal N_k(t,x;\theta_{\mathrm n}) + b_0$ is a DeepONet of the form \eqref{eq: deeponet}. The ansatz \eqref{eq: hard BC} satisfies the terminal condition by construction, that is, $\hat{\mathcal T}_\theta^{\mathrm{hbc}}[a](t_{\mathrm f},x) = \max\{r(x),a(x)\}$ for every $a\in\mathcal A_{\mathrm c}(\Omega',M,L)$ and $x\in \Omega$, eliminating the need to separately penalize boundary mismatches during training. 
Since $r$ and $a$ are uniformly Lipschitz continuous on $\Omega'$, the map $x \mapsto \max\{r(x),a(x)\}$ is differentiable almost everywhere by Rademacher's theorem, so that the spatial derivatives of $\hat{\mathcal T}_\theta^{\mathrm{hbc}}[a]$ required to evaluate the HJI residual exist at almost every $x\in\Omega$ and, with probability one, at every collocation point sampled from an absolutely continuous distribution.

\subsection{Universal Approximation} 
The next theorem establishes the universal approximation property of the RA-PI-DeepONet, showing that the architecture \eqref{eq: hard BC} is sufficiently expressive to approximate $\mathcal T$ to arbitrary accuracy on the compact class of avoid functions $\mathcal A_{\mathrm c}(\Omega',M,L)$.

\begin{theorem}
Let $\Omega \subset \mathbb R^n$ be a nonempty compact set, let $\Omega'$ be given by \eqref{eq: inflated set}, let $M,L>0$, let $\mathcal A_{\mathrm c}(\Omega',M,L)$ be given by \eqref{eq:compact_avoid_class}, and let $\sigma:\mathbb R \to \mathbb R$ be a continuous nonpolynomial function. Then, for every $\epsilon>0$, there exist positive integers $m,p$, sensor points $\{x_j\}_{j=1}^m \subset \Omega'$, a branch network $\mathcal B(\cdot;\theta_{\mathrm b}):\mathbb R^m\to\mathbb R^p$ and a trunk network $\mathcal N(\cdot,\cdot;\theta_{\mathrm n}):[t_0,t_{\mathrm f}]\times\mathbb R^n\to\mathbb R^p$ with activation function $\sigma$, and a bias $b_0\in\mathbb R$, such that the RA-PI-DeepONet $\hat{\mathcal T}^{\mathrm{hbc}}_\theta$ defined by \eqref{eq: hard BC}, with $\theta=(\theta_{\mathrm b},\theta_{\mathrm n},b_0)$, satisfies
\begin{equation}\label{eq: UAT bound}
\sup_{a\in \mathcal A_{\mathrm c}(\Omega',M,L)}\ \sup_{(t,x)\in [t_0,t_{\mathrm f}]\times\Omega} \big|\hat{\mathcal T}^{\mathrm{hbc}}_\theta[a](t,x)-\mathcal T[a](t,x)\big|<\epsilon.
\end{equation}
\end{theorem}
\begin{proof}
For every $a\in\mathcal A_{\mathrm c}(\Omega',M,L)$ and every $(t,x)\in[t_0,t_{\mathrm f}]\times\Omega$, define the shifted solution operator
$
\widetilde{\mathcal T}[a](t,x)
\triangleq
\mathcal T[a](t,x)-\max\{r(x),a(x)\}.
$
Next, for every $a_1, a_2\in\mathcal A_{\mathrm c}(\Omega',M,L)$, using
$
\big|\max\{r(x),a_1(x)\}-\max\{r(x),a_2(x)\}\big|
\leq
|a_1(x)-a_2(x)|
$
and \eqref{eq: local Lipschitz}, it follows that
$
\|\widetilde{\mathcal T}[a_1]-\widetilde{\mathcal T}[a_2]\|_{\infty,[t_0,t_{\mathrm f}]\times\Omega}
\leq
2\|a_1-a_2\|_{\infty,\Omega'}.
$
Hence, $\widetilde{\mathcal T}$ is Lipschitz continuous on $\mathcal A_{\mathrm c}(\Omega',M,L)$, and since $\widetilde{\mathcal T}[a]\in C([t_0,t_{\mathrm f}]\times\Omega)$ for every $a\in\mathcal A_{\mathrm c}(\Omega',M,L)$, the map $(a,t,x)\mapsto\widetilde{\mathcal T}[a](t,x)$ is continuous. Furthermore, the map $(a,t,x)\mapsto\widetilde{\mathcal T}[a](t,x)$ is uniformly continuous since $\mathcal A_{\mathrm c}(\Omega',M,L)\times[t_0,t_{\mathrm f}]\times\Omega$ is compact, and hence, by the terminal condition \eqref{thm_eq:boundary condition},
\begin{equation} \label{eq: shifted terminal}
\lim_{t\uparrow t_{\mathrm f}}
\sup_{a\in\mathcal A_{\mathrm c}(\Omega',M,L)}
\sup_{x\in\Omega}
\big|\widetilde{\mathcal T}[a](t,x)\big|
=0.
\end{equation}

Next, for $\delta>0$, define the regularized operator
$
Q_\delta[a](t,x)
\triangleq
\frac{\widetilde{\mathcal T}[a](t,x)}
{t_{\mathrm f}-t+\delta}.
$
Since, for every $a_1,a_2\in\mathcal A_{\mathrm c}(\Omega',M,L)$,
$
\|Q_\delta[a_1]-Q_\delta[a_2]\|_{\infty,[t_0,t_{\mathrm f}]\times\Omega}
\leq
\frac{2}{\delta}
\|a_1-a_2\|_{\infty,\Omega'},
$
$Q_\delta$ is Lipschitz continuous, and hence, by the generalized universal approximation theorem for operators \cite[Theorem 2]{lu2021learning}, for every $\eta>0$, there exists a DeepONet $\mathcal G_\theta$ such that
\begin{equation} \label{eq: Q UAT}
\sup_{a\in\mathcal A_{\mathrm c}(\Omega',M,L)}
\sup_{(t,x)\in[t_0,t_{\mathrm f}]\times\Omega}
\big|Q_\delta[a](t,x)-\mathcal G_\theta[a](t,x)\big|
<\eta.
\end{equation}

Next, define $ M_{\widetilde{\mathcal T}}
\triangleq
\sup_{a\in\mathcal A_{\mathrm c}(\Omega',M,L)}
\sup_{(t,x)\in[t_0,t_{\mathrm f}]\times\Omega}
\big|\widetilde{\mathcal T}[a](t,x)\big|.
$
Given $\epsilon>0$, by \eqref{eq: shifted terminal}, there exists $\rho>0$ such that
$
\big|\widetilde{\mathcal T}[a](t,x)\big|
<
\frac{\epsilon}{2}
$
for every $a\in\mathcal A_{\mathrm c}(\Omega',M,L)$ and $x\in\Omega$ whenever $t_{\mathrm f}-t<\rho$. 

Now, let $\delta>0$ be such that
$
M_{\widetilde{\mathcal T}}\frac{\delta}{\rho}
<
\frac{\epsilon}{2}.
$
If $t_{\mathrm f}-t<\rho$, then

\begin{align*}
\left|
\widetilde{\mathcal T}[a](t,x)
-(t_{\mathrm f}-t)Q_\delta[a](t,x)
\right|
&=
\big|\widetilde{\mathcal T}[a](t,x)\big|
\frac{\delta}{t_{\mathrm f}-t+\delta} \\
&\leq
\big|\widetilde{\mathcal T}[a](t,x)\big| \\ &
<
\frac{\epsilon}{2}.
\end{align*}

If $t_{\mathrm f}-t\geq\rho$, then
\begin{align*}
\left|
\widetilde{\mathcal T}[a](t,x)
-(t_{\mathrm f}-t)Q_\delta[a](t,x)
\right|
&=
\big|\widetilde{\mathcal T}[a](t,x)\big|
\frac{\delta}{t_{\mathrm f}-t+\delta}\\
&\leq
M_{\widetilde{\mathcal T}}\frac{\delta}{\rho} \\
& <
\frac{\epsilon}{2}.
\end{align*}
Thus,
\begin{gather} 
\sup_{a\in\mathcal A_{\mathrm c}(\Omega',M,L)}
\sup_{(t,x)\in[t_0,t_{\mathrm f}]\times\Omega}
\left|
\widetilde{\mathcal T}[a](t,x)
-(t_{\mathrm f}-t)Q_\delta[a](t,x)
\right| \nonumber \\
<
\frac{\epsilon}{2}.\label{eq: regularization bound}
\end{gather}

Now, let $\eta>0$ be such that
$
(t_{\mathrm f}-t_0)\eta<\frac{\epsilon}{2}.
$
Then, by \eqref{eq: Q UAT} and \eqref{eq: regularization bound}, for every $a\in\mathcal A_{\mathrm c}(\Omega',M,L)$ and $(t,x)\in[t_0,t_{\mathrm f}]\times\Omega$,
\begin{align*}
\big|
\mathcal T[a](t,x)
&-\hat{\mathcal T}^{\mathrm{hbc}}_\theta[a](t,x)
\big|
\\  \leq &
\big|
\widetilde{\mathcal T}[a](t,x)
-(t_{\mathrm f}-t)Q_\delta[a](t,x)
\big|\\
&+
(t_{\mathrm f}-t)
\big|
Q_\delta[a](t,x)-\mathcal G_\theta[a](t,x)
\big|\\
<&
\frac{\epsilon}{2}
+
(t_{\mathrm f}-t_0)\eta \\
< & \epsilon,
\end{align*}
which yields \eqref{eq: UAT bound}.
\frQED
\end{proof}
 
\subsection{Physics-Informed Training} 
To enable training of the RA-PI-DeepONet without precomputed value functions, we leverage the HJI variational inequality \eqref{eq: HJB} together with the boundary condition \eqref{thm_eq:boundary condition} as physics-based supervision. Specifically, given an avoid function $a\in\mathcal A$, for a function $V(t,x)$ differentiable at $(t,x)\in[t_0,t_{\mathrm f})\times\mathbb R^n$, define the HJI residual operator $\mathcal R_a$ by
\begin{align*}
\mathcal R_a[V] &(t,x) \triangleq  \max \Big\{ \frac{\partial V}{\partial t}(t,x) +\min\{0,  \\ &
\sup_{d \in  D}\inf_{u \in  U}
\frac{\partial V}{\partial x}(t,x)\, F(x,u,d)
\}, a(x) - V(t,x) \Big\}.
\end{align*}
By Theorem \ref{Thm: Nash equilibrium}, for every $a\in\mathcal A$, the value function $V_a=\mathcal T[a]$ is globally Lipschitz continuous and satisfies $\mathcal R_a[V_a](t,x)=0$ almost everywhere on $[t_0,t_{\mathrm f})\times\mathbb R^n$ and the terminal condition \eqref{thm_eq:boundary condition}.

For the training procedure, we further assume that $\Omega$ has nonempty interior. At every optimization iteration, we draw a fresh batch consisting of $N_{\mathrm a}$ avoid functions $\{a^{(i)}\}_{i=1}^{N_{\mathrm a}}\subset\mathcal A_{\mathrm c}(\Omega',M,L)$ from a user-specified distribution over $\mathcal A_{\mathrm c}(\Omega',M,L)$ and, for every $i\in\{1,\ldots,N_{\mathrm a}\}$, $N_{\mathrm r}$ collocation points $\{(t^{(i,j)},x^{(i,j)})\}_{j=1}^{N_{\mathrm r}}\subset[t_0,t_{\mathrm f})\times\Omega$ from an absolutely continuous distribution. The physics-informed empirical loss is then defined as
\begin{equation}\label{eq: physics loss}
\mathcal L(\theta)
\triangleq
\frac{1}{N_{\mathrm a}N_{\mathrm r}}
\sum_{i=1}^{N_{\mathrm a}}
\sum_{j=1}^{N_{\mathrm r}}
\left|
\mathcal R_{a^{(i)}}[\hat{\mathcal T}^{\mathrm{hbc}}_\theta[a^{(i)}]]
(t^{(i,j)},x^{(i,j)})
\right|^2.
\end{equation}

The HJI residual $\mathcal R_{a^{(i)}}[\hat{\mathcal T}^{\mathrm{hbc}}_\theta[a^{(i)}]](t,x)$ in \eqref{eq: physics loss} involves the partial derivatives $\partial_t\hat{\mathcal T}^{\mathrm{hbc}}_\theta[a^{(i)}]$ and $\partial_x\hat{\mathcal T}^{\mathrm{hbc}}_\theta[a^{(i)}]$, which are computed by reverse-mode automatic differentiation. 
Training amounts to determining parameters $\theta$  that minimize $\mathcal L(\theta)$ via a stochastic gradient-based optimizer. Since the loss \eqref{eq: physics loss} aggregates HJI residuals over a batch of distinct avoid functions $\{a^{(i)}\}_{i=1}^{N_{\mathrm a}}$, the trained RA-PI-DeepONet is encouraged to approximate $V_{a^{(i)}}$ simultaneously for every $i \in \{1,\ldots,N_{\mathrm a}\}$, and hence, to approximate the operator $\mathcal T$ over $\mathcal A_{\mathrm c}(\Omega',M,L)$. The training procedure is summarized in Algorithm \ref{alg: training}.

\begin{algorithm}
\caption{Training of RA-PI-DeepONet}
\label{alg: training}
\hspace*{\algorithmicindent} \textbf{Hyperparameters}: $N_{\mathrm a}, N_{\mathrm r}, \mathcal K$\\
\hspace*{\algorithmicindent} \textbf{Input}: Sensor points $\{x_j\}_{j=1}^m\subset\Omega'$ and sampling distributions for avoid functions on $\mathcal A_{\mathrm c}(\Omega',M,L)$ and collocation points on $[t_0,t_{\mathrm f})\times\Omega$\\
\hspace*{\algorithmicindent} \textbf{Output}: Trained RA-PI-DeepONet $\hat{\mathcal T}^{\mathrm{hbc}}_\theta$
\begin{algorithmic}[1]
    \State Initialize network parameters $\theta$
    \For{$k=1,\ldots,\mathcal K$}
        \State Sample $\{a^{(i)}\}_{i=1}^{N_{\mathrm a}}\subset\mathcal A_{\mathrm c}(\Omega',M,L)$
         \State Form the encoded inputs $\{\bar a^{(i)}\}_{i=1}^{N_{\mathrm a}}$ using \eqref{eq: encoded}
         \State Sample $\{(t^{(i,j)},x^{(i,j)})\}_{j=1}^{N_{\mathrm r}}\subset[t_0,t_{\mathrm f})\times\Omega$ for every $i\in\{1,\ldots,N_{\mathrm a}\}$
        \State Evaluate $\mathcal L(\theta)$ from \eqref{eq: physics loss}
        \State Update $\theta$ with one step of the stochastic gradient-based optimizer
    \EndFor
\end{algorithmic}
\end{algorithm}

Once the parameters $\theta$ are obtained by minimizing the empirical loss \eqref{eq: physics loss}, for every $a\in\mathcal A_{\mathrm c}(\Omega',M,L)$ and $(t,x)\in[t_0,t_{\mathrm f}]\times\Omega$, we consider the masked approximation
\begin{equation}\label{eq: masked}
\hat{\mathcal T}^{\mathrm m}_\theta[a](t,x) \triangleq
\begin{cases}
\max\{r(x),a(x)\}, & \text{if } r(x) \leq0,\\
\max\big\{\hat{\mathcal T}_\theta^{\mathrm{hbc}}[a](t,x),\ a(x)\big\}, & \text{otherwise}.
\end{cases}
\end{equation}

The reach-avoid set $\mathcal S(R,A(\tilde a))$,
where $\tilde a\in\mathcal A$ is any extension of
$a\in\mathcal A_{\mathrm c}(\Omega',M,L)$ to $\mathbb R^n$,
is then approximated on $\Omega$ by the
\textit{neural reach-avoid set}
\begin{equation*}
\hat{\mathcal S}_\theta(a) \triangleq
\big\{x\in\Omega:\hat{\mathcal T}^{\mathrm m}_\theta[a](t_0,x)\leq0\big\}.
\end{equation*}
By construction of \eqref{eq: masked},
$\hat{\mathcal S}_\theta(a)$ contains
$(R\setminus A(\tilde a))\cap\Omega$ and excludes
$A(\tilde a)\cap\Omega$ for every
$a\in\mathcal A_{\mathrm c}(\Omega',M,L)$,
independently of the network approximation error. Since evaluating $\hat{\mathcal T}^{\mathrm m}_\theta[a]$ requires only a forward pass through the trained networks, the neural reach-avoid set can be computed in real time whenever the avoid function changes, without re-solving the HJI variational inequality.

\section{Numerical Results} \label{sec: simulations}
In this section, we demonstrate the proposed framework on two reach-avoid differential games involving a planar underwater vehicle with a three-dimensional state space, for which grid-based level-set solvers remain tractable and provide a numerical baseline, and a fixed-wing UAV with a six-dimensional state space (3D-UAV), for which grid-based solvers are computationally intractable. The RA-PI-DeepONet models for both examples are implemented in PyTorch and trained on a single NVIDIA H100 GPU by minimizing the residual loss \eqref{eq: physics loss} under the hard boundary-condition ansatz \eqref{eq: hard BC}, and the neural reach-avoid sets are computed from the masked approximation \eqref{eq: masked}.

\subsection{Planar Underwater Vehicle} 

\begin{figure*}[!t]
    \centering
    \includegraphics[width=0.72\linewidth]{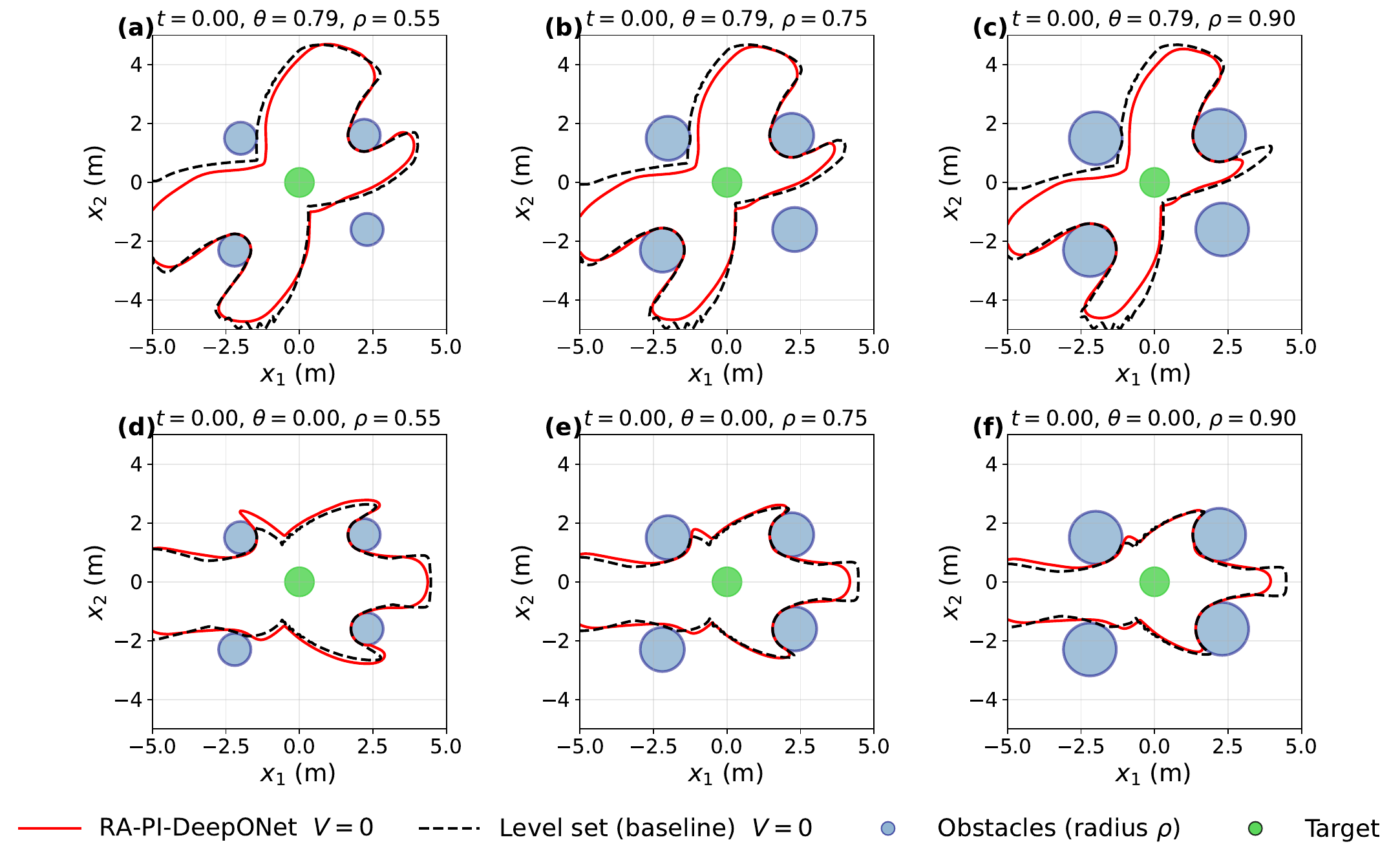}
    \caption{Comparison of the trained RA-PI-DeepONet with the level-set baseline for the underwater-vehicle reach-avoid game. Each panel shows the zero level set of the value function at $t=0$ computed by the RA-PI-DeepONet (solid red) and by the level-set method (dashed black), for headings $\theta = \pi/4$ (top row) and $\theta = 0$ (bottom row), and obstacle radii $\rho = 0.55$, $\rho = 0.75$, and $\rho = 0.90$ (left to right). The blue disks depict the obstacles and the green disk depicts the target set.}
    \label{fig: dubin operator baseline}
\end{figure*}
We consider the planar underwater vehicle described in \cite{margellos2011hamilton}, which navigates at a constant depth subject to a constant ambient current and a bounded adversarial cross-current, with the state $x = [x_1,\ x_2,\ \theta]^{\mathrm T}$ comprising the planar position $(x_1,x_2)$ and the heading $\theta$. The guidance law comprises the commanded speed and heading rate, and the planar kinematics and corresponding extremal inputs are given in \cite{margellos2011hamilton}. The value function is approximated on $\Omega \triangleq [-5,\,5]^2 \times [-\pi,\,\pi]$ over the horizon $[t_0,\,t_{\mathrm f}] = [0,\,3]$. The reach set, interpreted as the target set for the vehicle, is defined by a disk of radius $0.5$ centered at the origin in the position subspace, encoded by the reach function $r(x) = \|[x_1,\ x_2]^{\mathrm T}\| - 0.5$, and the avoid functions are drawn from the one-parameter family
\begin{equation*}
a(x;\rho) \triangleq \rho - \min_{j \in \{1,\ldots,4\}} \big\| [x_1,\ x_2]^{\mathrm T} - c_j \big\|,
\end{equation*}
encoding four disjoint circular obstacles with centers $c_j$, $j\in\{1,\ldots,4\}$, in the position subspace and a shared radius $\rho$ varying in $[0.5,\,1]$. Here and in the 3D-UAV example below, the reach and avoid functions are unbounded on $\mathbb R^n$ and are truncated outside a sufficiently large ball containing $\Omega'$, which affects neither their values along solutions originating in $\Omega$ nor the associated value functions on $[t_0,\,t_{\mathrm f}]\times\Omega$. We select a single sensor point $\xi\in\Omega'$ whose position coincides with one of the obstacle centers. Hence, $a(\xi;\rho)=\rho$, and the encoded input \eqref{eq: encoded} with $m=1$ reduces to $\bar a=\rho$.

To assess the accuracy of the trained RA-PI-DeepONet, we compare the neural reach-avoid sets with those obtained using a classical level-set method for the HJI variational inequality \eqref{eq: HJB} \cite{mitchell2002application,mitchell2005time}, computed on a $221 \times 221 \times 72$ grid that is periodic in the heading $\theta$.
For consistency, the masking operation in \eqref{eq: masked} is also applied to the level-set solutions after every time step. 
Note that each level-set solution requires approximately $15\,\mathrm s$ on a desktop CPU and must be recomputed whenever the avoid function changes, whereas the trained RA-PI-DeepONet evaluates a $221 \times 221$ slice of the value function for any $\rho \in [0.5,\,1]$ in approximately $0.16\,\mathrm s$ on the same hardware.

Figure~\ref{fig: dubin operator baseline} compares the zero level sets of the RA-PI-DeepONet and level-set approximations at $t = 0$ for headings $\theta \in \{0,\ \pi/4\}$ and obstacle radii $\rho \in \{0.55,\ 0.75,\ 0.90\}$. In all six slices, the boundaries of the neural reach-avoid sets are in close agreement with the level-set baseline and capture the deformation of the reach-avoid sets as $\rho$ increases, although all the neural reach-avoid sets are obtained from a single trained network. In particular, over the six slices of Figure~\ref{fig: dubin operator baseline}, the two approximations agree in sign on $94.3$--$98.0\%$ of the grid points, with an intersection over union of the corresponding zero sublevel sets of $81$--$92\%$ and a root-mean-square discrepancy of $0.17$--$0.20$ for values ranging over approximately $[-2.0,\ 3.7]$, whereas over the full three-dimensional grid, the sign agreement is $92.0$--$92.4\%$ and the root-mean-square discrepancy is approximately $0.6$, concentrated near $\theta = \pm\pi$ since the trunk network receives the raw heading angle and hence does not enforce periodicity in $\theta$.

\begin{figure}[!t]
    \centering
    \includegraphics[width=\columnwidth]{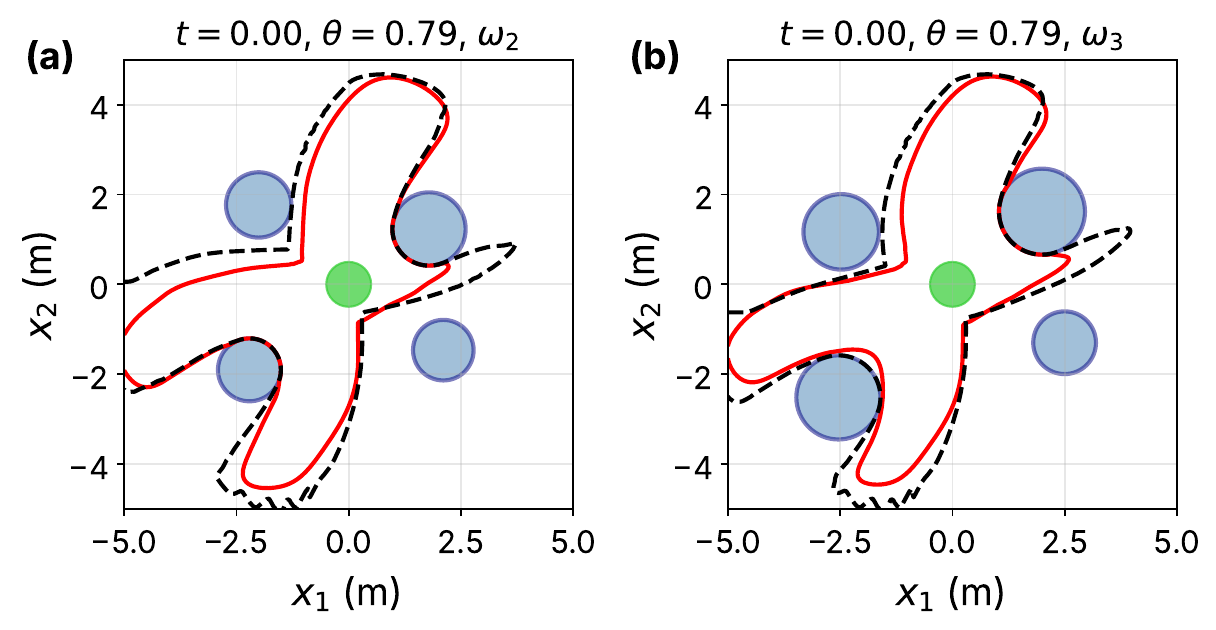}
    \caption{Zero level sets of the value function at $t=0$ and $\theta = \pi/4$ computed by the RA-PI-DeepONet (solid red) and by the level-set method (dashed black) for the underwater-vehicle reach-avoid game with obstacles (blue) of varying centers and radii, for the configurations $\omega_2$ (a) and $\omega_3$ (b), where the green disk depicts the target set.}
    \label{fig: dubin operator pos baseline}
\end{figure}

To further demonstrate the flexibility of the RA-PI-DeepONet, we also consider obstacles whose centers $c_j = \bar c_j + \delta_j$, with $\delta_j \in [-0.5,\,0.5]^2$, and radii $\rho_j \in [0.5,\,1]$, $j\in\{1,\ldots,4\}$, vary independently, where $\bar c_j$ denotes the nominal center used above, and the branch network receives the parameter vector $\omega \triangleq (c_1,\rho_1,\ldots,c_4,\rho_4)$ determining the avoid function in place of \eqref{eq: encoded}. A separate RA-PI-DeepONet is trained for this family using the same procedure. Figure~\ref{fig: dubin operator pos baseline} compares the RA-PI-DeepONet and level-set approximations at $t = 0$ and $\theta = \pi/4$ for two configurations $\omega_2$ and $\omega_3$ drawn at random from the training distribution. In both slices, the neural reach-avoid sets closely follow the level-set baseline and adapt to the individual positions and sizes of the obstacles, although both of them are obtained from a single trained network, with a sign agreement of $93.1$--$94.0\%$, an intersection over union of $78$--$80\%$, and a root-mean-square discrepancy of $0.17$--$0.21$ over the two slices, and a sign agreement of $90.9$--$91.6\%$ over the full three-dimensional grid.

\subsection{3D-UAV} 

\begin{figure*}[!t]
    \centering
    \includegraphics[width=0.85\linewidth]{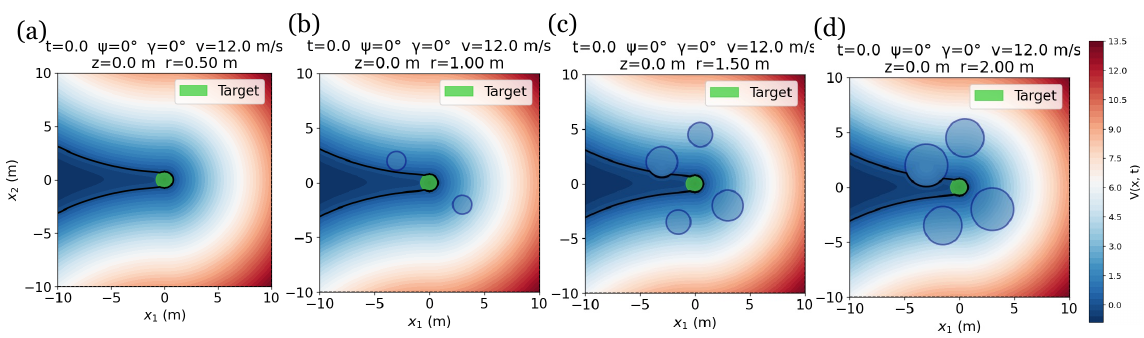}
    \caption{Neural reach-avoid sets computed by RA-PI-DeepONet for the 3D-UAV reach-avoid game, for which grid-based level-set solvers are computationally intractable. Each panel shows the value function at $t=0$ (color) and its zero level set (black) in the $(x_1,x_2)$ plane, with the remaining states fixed at $x_3 = 0$, $\psi = 0$, $\gamma = 0$, and $v = 12$, for obstacle radii $\rho = 0.5$ (a), $\rho = 1.0$ (b), $\rho = 1.5$ (c), and $\rho = 2.0$ (d). The blue disks depict the $x_3 = 0$ cross-sections of the spherical obstacles, which are empty when the centers lie more than $\rho$ from this plane, as in (a), and the green disk depicts the cross-section of the target set.}
    \label{fig: uav operator}
\end{figure*}

We consider a fixed-wing UAV whose six-dimensional state $x = [x_1,\ x_2,\ x_3,\ \psi,\ \gamma,\ v]^{\mathrm T}$ comprises the position $p \triangleq [x_1,\ x_2,\ x_3]^{\mathrm T} \in \mathbb R^3$, the heading angle $\psi$, the flight path angle $\gamma$, and the airspeed $v \geq 0$. The guidance law comprises the commanded heading, flight path angle, and airspeed rates. The nominal fixed-wing kinematics are adopted from \cite{kokolakis2026fixed}, while a bounded adversarial disturbance acts on the position kinematics. The value function is approximated on $\Omega \triangleq [-10,\,10]^2 \times [-5,\,5] \times [-\pi,\,\pi] \times [-\pi/3,\,\pi/3] \times [5,\,20]$ over the horizon $[t_0,\,t_{\mathrm f}] = [0,\,5]$. The reach set is defined by a ball of radius $0.8$ centered at the origin in the position subspace, encoded by the reach function $r(x) = \|[x_1,\ x_2,\ x_3]^{\mathrm T}\| - 0.8$. The avoid functions are drawn from the one-parameter family
\begin{equation*}
a(x;\rho) \triangleq \rho - \min_{j \in \{1,\ldots,4\}} \|p-c_j\|,
\end{equation*}
encoding four disjoint spherical obstacles with fixed centers $c_j \in \mathbb R^3$, $j\in\{1,\ldots,4\}$, in the position subspace and a shared radius $\rho$ varying in $[0.5,\,2]$. As in the underwater-vehicle example, the encoded input \eqref{eq: encoded} with a single sensor point at one of the obstacle centers reduces to $\bar a=\rho$.

Since the HJI variational inequality \eqref{eq: HJB} is now posed over a six-dimensional state space, grid-based level-set solvers are computationally intractable, and hence, we present only the RA-PI-DeepONet results.

Figure~\ref{fig: uav operator} shows the resulting neural reach-avoid sets at $t = 0$ in the $(x_1,x_2)$ plane, with the remaining states fixed at $x_3 = 0$, $\psi = 0$, $\gamma = 0$, and $v = 12$, for obstacle radii $\rho \in \{0.5,\ 1.0,\ 1.5,\ 2.0\}$. As $\rho$ increases, the neural reach-avoid set contracts and deforms around the growing obstacles. Note that all four panels are obtained from a single trained network by varying the branch input $\rho$ at inference time.

\section{Conclusion}
In this paper, we introduced RA-PI-DeepONet, a physics-informed neural operator framework for approximating the solution operator associated with the reach-avoid game without requiring paired input-output training data. We established that the solution operator mapping an avoid function to the corresponding value function is Lipschitz continuous and, by the universal approximation theorem for operators, can be approximated by RA-PI-DeepONet to arbitrary accuracy on compact sets of avoid functions. Trained offline by minimizing the HJI variational inequality residual with the boundary condition enforced by construction, RA-PI-DeepONet enables real-time inference of the value function for unseen avoid functions, eliminating the need to repeatedly solve the HJI variational inequality whenever the avoid function changes. Future research will explore extensions of this framework to discrete-time reach-avoid games.
\balance
\bibliographystyle{IEEEtran}
\bibliography{references}
\end{document}